\documentclass[pra,twocolumn]{revtex4}
\usepackage{amsmath}
\usepackage{amssymb}
\usepackage{graphicx}

\begin{document}

\title{Quantum state transmission in a cavity array via two-photon exchange}
\author{Yu-Li Dong, Shi-Qun Zhu, and Wen-Long You}
\affiliation{School of Physical Science and Technology, Soochow University, Suzhou, Jiangsu 215006, P. R. China}

\begin{abstract}
The dynamical behavior of a coupled cavity array is investigated when each
cavity contains a three-level atom. For the uniform and staggered
intercavity hopping, the whole system Hamiltonian can be analytically
diagonalized in the subspace of single-atom excitation. The quantum state
transfer along the cavities is analyzed in detail for distinct regimes of
parameters, and some interesting phenomena including binary transmission,
selective localization of the excitation population are revealed. We
demonstrate that the uniform coupling is more suitable for the quantum state
transfer. It is shown that the initial state of polariton located in the
first cavity is crucial to the transmission fidelity, and the local
entanglement depresses the state transfer probability. Exploiting the
metastable state, the distance of the quantum state transfer can be much
longer than that of Jaynes-Cummings-Hubbard model. A higher transmission
probability and longer distance can be achieved by employing a class of
initial encodings and final decodings.

PACS number: 42.50.Pq, 05.60.Gg, 03.67.Hk
\end{abstract}

\maketitle

\section{INTRODUCTION}

It is an important task in quantum information processing when a
quantum state is transferred from one location to another. It refers to not
only quantum communication \cite{zeilinger}, but also large-scale quantum
computing \cite{lloyd}. Different protocols are proposed to realize quantum
state transfer (QST). For example, based on measurement and reconstruction,
an unknown quantum state can be faithfully transferred to an arbitrarily
distant location via quantum teleportation \cite{ben}. During the quantum
teleportation, the measured result instead of the quantum state is
transferred by a classical communication channel. Alternatively, a quantum
state can be sent through a channel directly. There are two types of such
state transfer which depends on the transfer distance. For long distance
communication, the information encoded in photons can be transferred by an
optical fiber \cite{cirac97,boozer}, which has been widely used in quantum
communication and cryptography applications \cite{muller,ursin}. Compared
with the long distance communication, a kind of short distance communication
which can be used between adjacent quantum processors was recently
introduced \cite{bose03}. Using an unmodulated spin chain as a channel, the
state transmits from one end of the chain to another with some fidelity.
Subsequently, many schemes are proposed to gain higher even perfect transfer
fidelity \cite{chr,chr05,ross09,linden2004,zou,bish,lukin11}.

On account of the possibility of individual addressing, an array of coupled
cavities is probably a promising candidate for simulating spin chains. The
coupled cavities can be realized in various physical systems \cite{kimble},
such as photonic crystals \cite{hen}, superconducting resonators \cite{scho}%
, and cavity-fiber-cavity system \cite{sera}. So far, the Heisenberg chains
of spin $1/2$ \cite{har2007}, of any high spin \cite{cho}, and with
next-nearest-neighbor interactions \cite{chen} are simulated in the coupled
cavities. In these simulation processes, the cavity field is ingeniously
removed, and the same procedure occurs when the QST is investigated in Ref.
\cite{li09}. However, the array of coupled cavities, a hybrid system
combining the spinor atom and photons, would provide richer phenomena than
the pure spin chains or Bose-Hubbard model \cite%
{hartmann2006,greentre2006,mak2008}. For this reason, the temporal dynamics
\cite{kim08,mak,ciccarello} and the binary transmission \cite{zhang10,yang11}%
, i.e., the QST including both components in one-dimensional (1D) coupled
cavities, are analyzed in detail.

The binary transmission mentioned above is restricted in single-excitation
subspace, i.e., only one photon transfers along the cavities in cascades,
and usually the transfer distant extensively depends on the lifetime of
atomic excitation state. Quite recently, the two-photon propagation in
waveguide has been attracting increasing interest from many researchers,
such as two-photon transport in 1D waveguide coupled to a two-level emitter
\cite{fan07}, a three-level emitter \cite{roy}, and a nonlinear cavity \cite%
{law}. Compared with one-photon electric-dipole E1 transition, the
two-photon 2E1 transition that emits two photons simultaneously is a
second-order transition \cite{mokler}. Therefore, the metastable state
decaying via 2E1 transition has remarkable longer lifetime than the excited
state decaying via E1 transition. The typical example is the lifetime for
the $2^{2}S_{1/2}$ state in hydrogen is of the order of fractions of a
second, which is several orders longer than that of single-photon excitation
state. The long lifetime of the metastable state could provide sufficient
operation time to transfer information. More recently, Law \textit{et al.}
reveal that the photons trend to bound together \cite{law2011} in the
two-excitation subspace when the ratio of vacuum Rabi frequency to the
tunneling rate between cavities exceeds a critical value. Motivated by these
aspects, it is desired to explore the short distance binary transmission of
the model that bounding two photons as a quasiparticle \cite{alex2010,alex}.

In this work, we investigate the binary transmission in 1D array of coupled
cavities, each of which contains a three-level atom. By adiabatically
eliminating the intermediate state, the individual cavity can be described
by the two-photon Jaynes-Cummings (JC) model with a metastable state and a
ground state. In the single-atom excitation subspace, the system Hamiltonian
is able to be diagonalized exactly. It allows us to focus on the binary
transmission with the explicit expression of the excitation population. We
consider the uniform and staggered intercavity hopping in different
parameters regimes, especially in the strong coupling and strong hopping
limits. With the uniform hopping, we find that the initial state of
polariton located in the first cavity is essential to the transmission
fidelity. When the polariton is a pure atomic state, an oscillation behavior
in the envelopment of fidelity occurs and the state transfer probability
increases obviously. On resonance, we reveal the optimal transmission time
is mainly governed by the intercavity hopping strength and the system size.
We also show that the exploiting of the metastable state overcomes the
deficit of short transfer distance during the lifetime of excitation state
\cite{mak}. With the staggered hopping, the excitation population trends to
be localized in the first several cavities and slows down the transfer
speed. This phenomenon becomes more distinct as the distortion of the
hopping strength enlarges. In the free evolution of the system, the
transmission fidelity drops quickly as the size of the array increases. How
to enhance the fidelity of the transmission is important, since the fidelity
of QST is expected to be better than that by using straightforward classical
communication. We finally demonstrate that a class of initial encoding and
final decoding process, which is currently used in spin chains \cite%
{zou,bish}, can also greatly improve the performance of binary transmission.

The paper is organized as follows. In Sec. II, we discuss the dynamics of an
array of coupled cavities with the uniform coupling. In Sec. III, we
consider the general situation with staggered coupling between the adjacent
cavities. Section IV gives an effective method to improve the fidelity of
the quantum state transfer by encoding and decoding limited qubits in strong
coupling regime. In Sec. V, we give a brief conclusion to the paper.

\section{THEORY MODEL}

We consider an array of coupled cavities, each of which contains a
three-level atom in cascade configuration. The individual cavity is tuned to
two-photon resonance with the metastable state $\left\vert e\right\rangle $
and the ground state $\left\vert g\right\rangle $, and these two states are
coupled to an intermediate state $\left\vert i\right\rangle $ with a single
mode of the electromagnetic field with coupling strengths $g_{1}$ and $g_{2}$
respectively. The intermediate level is detuned by the amount $\delta $ from
the average of the energies of the $\left\vert e\right\rangle $ and $%
\left\vert g\right\rangle $. In the large detuning condition $|\delta
|>>g_{1},g_{2}$, the intermediate state can be adiabatically eliminated, and
the Hamiltonian of one individual cavity can be written as a two-photon JC
model \cite{buzek,alex1998} (here $\hbar =1$)
\begin{equation}
\mathcal{H}^{(i)}=\omega _{a}\sigma _{ee}^{(i)}+\omega _{c}a_{i}^{\dagger
}a_{i}+\lambda \left( \sigma _{eg}^{(i)}a_{i}^{2}+\sigma
_{ge}^{(i)}a_{i}^{\dagger 2}\right) ,  \label{singlecavity}
\end{equation}%
where $a_{i}$ ($a_{i}^{\dagger }$) is the photonic annihilation (creation)
operator, and $\sigma _{ge}^{(i)}=\left\vert g\right\rangle
^{(i)(i)}\left\langle e\right\vert $ ($\sigma _{eg}^{(i)}=\left\vert
e\right\rangle ^{(i)(i)}\left\langle g\right\vert $) is atomic transition
operator in the $i$th cavity, $\omega _{c}$ is the resonance frequency of
the cavity and $\omega _{a}$ is the effective energy of atom. The metastable
state $\left\vert e\right\rangle $ and the ground state $\left\vert
g\right\rangle $ is coupled by a two-photon process with strength $\lambda $%
, which is given as $\lambda =g_{1}g_{2}/\delta $. The two-photon JC model
has been experimentally realized in high-Q superconducting microwave cavity
\cite{brune1987}.

In Hamiltonian (\ref{singlecavity}), the operator $Q^{(i)}=a_{i}^{\dagger
}a_{i}+$ $\sigma _{ee}^{(i)}-\sigma _{gg}^{(i)}$ commutes with $\mathcal{H}%
^{(i)}$, implying that the total number of atomic inversion and photonic
excitation is conserved. The two-photon JC Hamiltonian can be diagonalized
analytically. Spanned by the subspace $\left\{ \left\vert e,n-2\right\rangle
^{(i)},\left\vert g,n\right\rangle ^{(i)}\right\} $ ($n$ $\geq $ $2$), where
$n$ is the number of photons, the Hamiltonian can be written by the
polaritonic state as
\begin{equation}
\mathcal{H}_{n}^{(i)}=\left(
\begin{array}{cc}
\omega _{a}+n\omega _{c}-2\omega _{c} & \lambda \sqrt{n\left( n-1\right) }
\\
\lambda \sqrt{n\left( n-1\right) } & n\omega _{c}%
\end{array}%
\right).
\end{equation}

The eigenvalues and eigenfunctions of $\mathcal{H}_{n}^{(i)}$ can be written
in terms of the dressed-state representation%
\begin{eqnarray}
\left\vert \pm ,n\right\rangle ^{(i)} &=&\chi ^{\pm }(\theta _{n})\left\vert
g,n\right\rangle ^{(i)}\pm \chi ^{\mp }(\theta _{n})\left\vert
e,n-2\right\rangle ^{(i)},  \label{dressstate} \\
E_{\pm ,n}^{(i)} &=&\frac{\Delta }{2}+n\omega _{c}\pm 2\chi _{n},\quad
\left( n\geqslant 2\right)
\end{eqnarray}%
where $\tan \theta _{n}={2\lambda \sqrt{n(1+n)}}/(\Delta +2\chi _{n})$, $%
\chi ^{+}\left( x\right) =\sin x$, $\chi ^{-}\left( x\right) =\cos x$, $\chi
_{n}=\sqrt{n(n-1)\lambda ^{2}+\Delta ^{2}/4}$, and the detuning $\Delta
=\omega _{a}-2\omega _{c}$.

The $N$ cavities linked via two-photon exchange are described by the
Hamiltonian
\begin{equation}
\mathcal{H}=\sum\limits_{i=1}^{N}\mathcal{H}^{(i)}+\xi
\sum\limits_{i,j=1}^{N}A_{ij}\left( a_{i}^{\dagger
2}a_{j}^{2}+a_{i}^{2}a_{j}^{\dagger 2}\right) ,  \label{mainHamiltonian}
\end{equation}%
where $\xi $ is the intercavity coupling of double photons and $A$ is the
adjacency matrix of the connectivity graph. Here the photons only hop
between the nearest-neighbor cavities, so the matrix element is given by $%
A_{ij}=\delta _{i,j\pm 1}$.

Since the hopping between cavities does not change the number of photons,
the operator $Q=\sum\nolimits_{i}Q^{(i)}$ commutes with the Hamiltonian $%
\mathcal{H}$. For simplicity, the analysis is restricted to the case of $%
Q=2-N$, i.e., the system contains only single-atom excitation or two-photon
excitation. Here it is assumed that all the cavities are equal. Then we
extract the site indexes from the atomic and field operators, and define
them as a set of bare bases which form a complete Hilbert space. The system
bases thus can be written as $\left\vert M\right\rangle \otimes \left\vert
g,2\right\rangle $ and $\left\vert M\right\rangle \otimes \left\vert
e,0\right\rangle $ $(M=1,2,...,N)$, which denote that there are two photons
or an atomic excitation at cavity $M$. In these restricted bases, the
Hamiltonian (\ref{mainHamiltonian}) can be represented as
\begin{equation}
\mathcal{H}_{\mathrm{res}}=\frac{\Delta }{2}I_{N}\otimes Z+\sqrt{2}\lambda
I_{N}\otimes X+2\xi A\otimes \frac{I_{2}-Z}{2},  \label{ham}
\end{equation}%
where $I_{m}$ is the $m\times m$ identity matrix and $X$, $Z$ are the pauli
operators acting on the certain site of cavity. If the open chain of the
coupled cavity is considered, i.e., the hard-wall boundary conditions are
assumed, the eigenvectors of the adjacency matrix $A$ are \cite{chr,mak} $%
\left\vert m\right\rangle =\sqrt{\frac{2}{N+1}}\sum\nolimits_{M}\sin \left(
\frac{mM\pi }{N+1}\right) \left\vert M\right\rangle $, with corresponding
eigenvalues $E_{m}=-2\cos \frac{m\pi }{N+1}$ for all $m=1,2,...,N$. For a 1D
chain of cavities, the Hamiltonian $\mathcal{H}_{\mathrm{res}}$ can be
expressed in the bases $\left\{ \left\vert m\right\rangle \otimes \left\vert
e,0\right\rangle ,\left\vert m\right\rangle \otimes \left\vert
g,2\right\rangle \right\} $ as block diagonal matrixes, in which the $m$th
block appears as
\begin{equation}
\mathcal{H}_{\mathrm{res}}(m)=\left(
\begin{array}{cc}
\frac{\Delta }{2} & \sqrt{2}\lambda  \\
\sqrt{2}\lambda  & -\frac{\Delta }{2}-4\xi \cos \frac{m\pi }{N+1}%
\end{array}%
\right) .
\end{equation}%
\begin{figure}[tbp]
\centering\includegraphics[scale=0.3]{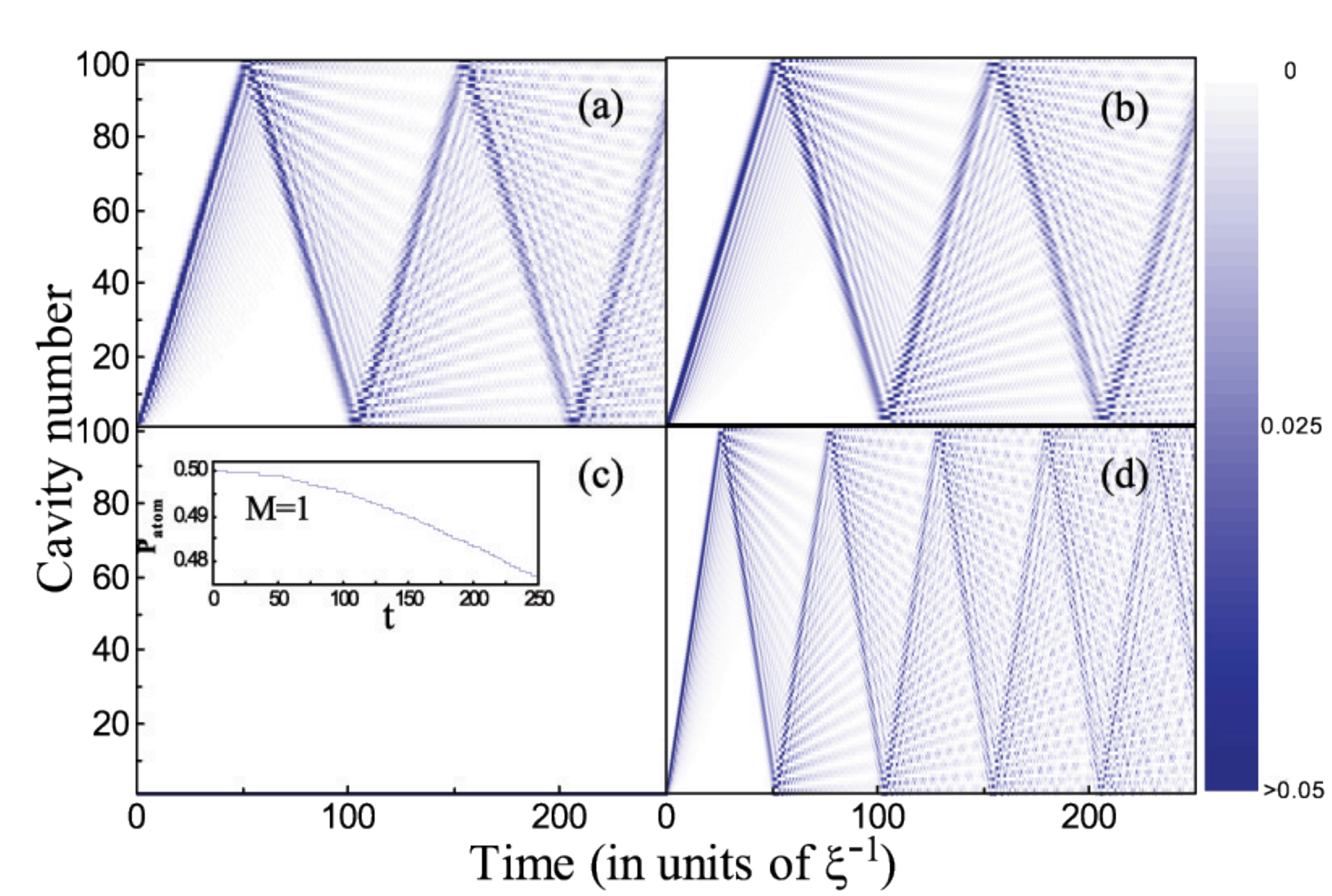} \caption{(Color
online) The probability of the certain cavity as a function of time
with the length of the chain $N=101$ and on resonance $\Delta =0$
for the initial state $\protect\beta =\protect\pi /4$. The upper two
plots show the population of atom (a) and photon (b) in the strong
J-C coupling regime ($\protect\lambda =200\protect\xi $), and the
lower two plots are the
population of atom (c) and photon (d) in the strong hopping regime ($\protect%
\lambda =\protect\xi /200$). The inset of (c) shows the population of the
atom in the first cavity slowly decreases with time. }
\label{population}
\end{figure}

Then the block matrix can be diagonalized, the eigenvalues are $E_{\pm
}^{m}=-2\xi \cos \left( \frac{m\pi }{N+1}\right) \pm \sqrt{\left[ \frac{%
\Delta }{2}+2\xi \cos \left( \frac{m\pi }{N+1}\right) \right] ^{2}+2\lambda
^{2}}$, and the corresponding eigenvectors are $\left\vert m,\pm
\right\rangle =\left\vert m\right\rangle \otimes \left\vert \pm
\right\rangle $, where the dressed state $\left\vert \pm \right\rangle
=[\left( \Delta -2E_{\mp }^{m}\right) \left\vert e,0\right\rangle +2\sqrt{2}%
\lambda \left\vert g,2\right\rangle ]/\sqrt{\left( \Delta -2E_{\mp
}^{m}\right) ^{2}+8\lambda ^{2}}$. With these analytical expressions, the
time evolution of arbitrary $N$ cavities can be straightforwardly derived.
In the following, the quantum state transfer is studied in such model. It
should be emphasized that there exist two channels, either atomic or
photonic, to transfer information, which is different from state transfer in
a pure spin chain. In this model, the sender and receiver can select which
qubit is encoded and decoded. The ground state of the system $\left\vert
g,0\right\rangle ^{\otimes N}$ with a zero eigenenergy will not involve any
evolution by the Hamiltonian $\mathcal{H}$. So only the propagation of the
excitation state for atom and photons needs to be considered.
\begin{figure}[tbp]
\centering\includegraphics[scale=0.30]{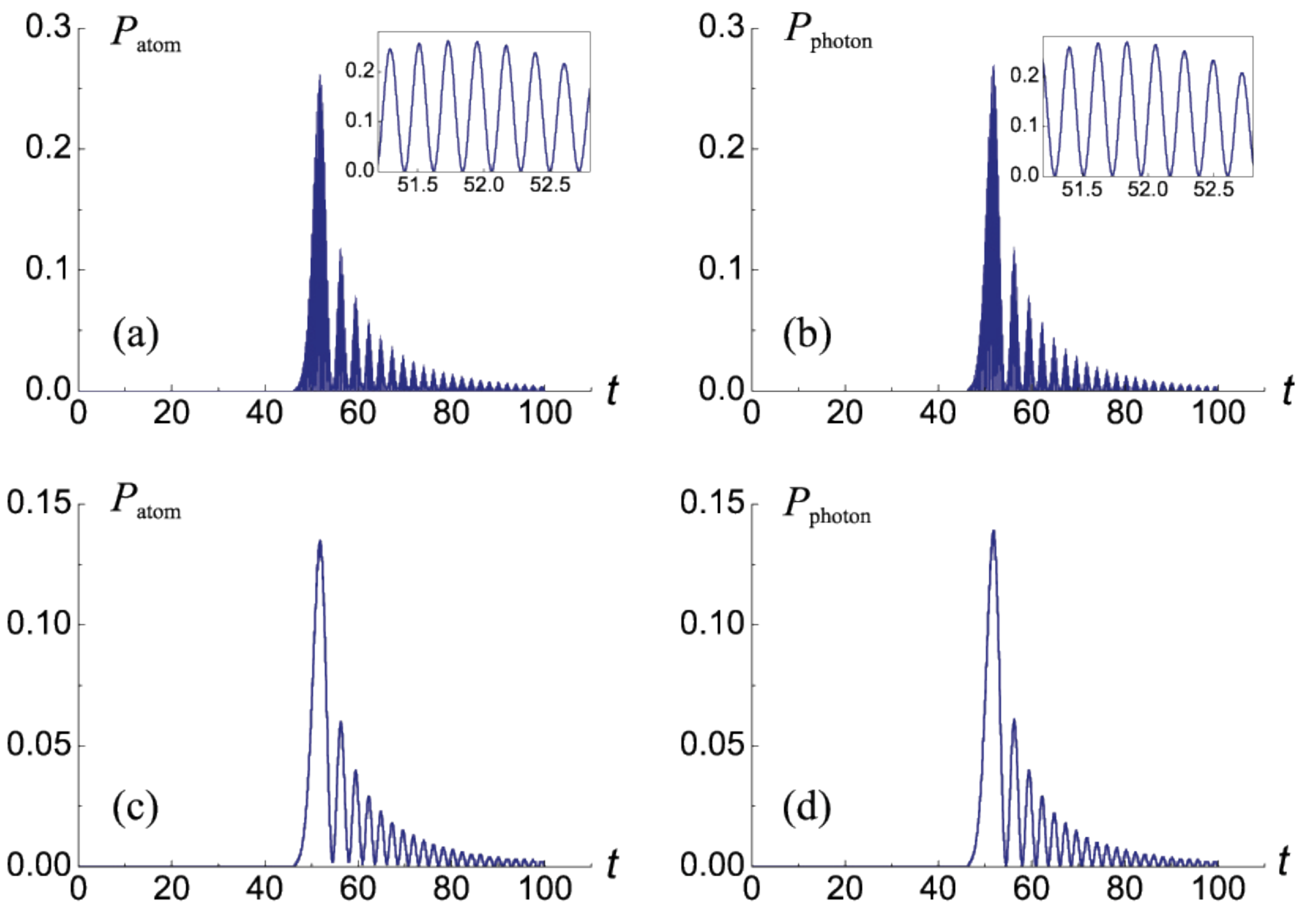} \caption{(Color
online) Dynamics of the transfer probabilities of atom and photon
respectively in strong coupling limit. The system parameters are $\
\Delta =0,\protect\lambda =10,\protect\xi =1,N=100$ for (a) and (b) $\protect%
\beta =\protect\pi /2$, (c) and (d) $\protect\beta =\protect\pi /4$. The
insets of (a) and (b) show the oscillation behavior\ in the envelopment.}
\label{coh}
\end{figure}

The influence of the initial state located in the first cavity is
investigated. Supposing that the initial state is an entangled state in the
first cavity
\begin{equation}
\left\vert \varphi (t=0)\right\rangle =\left\vert M=1\right\rangle \otimes
\left( \cos \beta \left\vert g,2\right\rangle +\sin \beta \left\vert
e,0\right\rangle \right) ,  \label{ini}
\end{equation}%
the state at time $t$ is given by $\left\vert \varphi (t)\right\rangle
=U(t)\left\vert \varphi (0)\right\rangle $, where $U(t)=e^{-i\mathcal{H}t}$
is the evolution operator of the whole system. Hence one has
\begin{equation}
\left\vert \varphi (t)\right\rangle =\sum\limits_{M,m=1}^{N}\left\vert
M\right\rangle \otimes \left[ f_{M,m}^{+}\left( t\right) \left\vert
+\right\rangle +f_{M,m}^{-}\left( t\right) \left\vert -\right\rangle \right]
,
\end{equation}%
where the probability amplitudes are given by
\begin{eqnarray}
f_{M,m}^{\pm }\left( t\right)  &=&\frac{2}{N+1}e^{-iE_{\pm }^{m}t}\chi ^{\pm
}\left( \beta -\alpha _{m}\right)   \notag \\
&&\times \sin \left( \frac{m\pi }{N+1}\right) \sin \left( \frac{mM\pi }{N+1}%
\right) ,
\end{eqnarray}%
with $\tan \alpha _{m}=-\{[\left( \Delta -2E_{+}^{m}\right) ^{2}+8\lambda
^{2}]/[\left( \Delta -2E_{-}^{m}\right) ^{2}+8\lambda ^{2}]\}^{1/2}$. Thus,
the probabilities of occupation for atom and photons at the $s$th cavity are
given by
\begin{equation}
P_{\mathrm{atom/photon,s}}=\left\vert \sum\limits_{m=1}^{N}\left[
f_{s,m}^{\pm }\chi ^{-}(\alpha _{m})\pm f_{s,m}^{\mp }\chi ^{+}\left( \alpha
_{m}\right) \right] \right\vert ^{2}.
\end{equation}%
It is assumed that the system is on resonant condition, i.e., $\Delta =0$.
If the coupling in each cavity dominates the evolution, i.e., in the so
called strong coupling limit $\lambda >>\xi $, the Hamiltonian becomes$\
\mathcal{H}_{\mathrm{res}}=\xi A\otimes I_{2}$ which has ignored the
fast-rotating terms in the interaction picture. Such Hamiltonian mimics two
Heisenberg spin chains \cite{kay2008} in atomic and photonic channels
respectively. Then the information transferred from sender to receiver can
be encoded in either channel with speed $\xi $, and the selectivity of
channels may be useful in binary communication. Compared with the Ref. \cite%
{mak}, due to the bounding effect of photon discussed here, the time for the
quantum state transfer is reduced by half given the both system have the
same hopping strength (see Fig. \ref{population} (a) and (b)). In other
limit, that the hopping dominates the evolution, i.e., $\xi >>\lambda $, the
Hamiltonian reduces to $\mathcal{H}_{\mathrm{res}}=2\xi A\otimes \left(
I_{2}-Z\right) /2$. Governed by such Hamiltonian, only the photon is
delocalized on the whole coupled cavity array with double speed $2\xi $
(Fig. \ref{population} (d)), whereas the atomic population is trapped in the
first cavity without dispersion (Fig. \ref{population} (c) and the inset in
it).

In the following of this section, we focus our interest on the binary
transmission in the strong coupling limit. We find that the temporal
evolution of the state vector sensitively depends on the initial state of
the system. When $\beta =\pi /2$, i.e., the initial localized excitation is
purely atomic, Figs. \ref{coh} (a) and (b) show that both probabilities
oscillate in each envelopment and have the same transmission behavior. When $%
\beta =\pi /4$, the initial state is the maximum entangled state localized
in the first cavity. Meanwhile $\alpha _{m}\simeq -\pi /4$ in the strong
coupling limit on resonance. Then the transfer probabilities can be
expressed simply as $P_{\mathrm{atom/photon,N}}\simeq 2\left\vert
\sum\nolimits_{m}e^{-iE_{+}^{m}t}\sin \left( \frac{m\pi }{N+1}\right) \sin
\left( \frac{mN\pi }{N+1}\right) \right\vert ^{2}/\left( N+1\right) ^{2}$.
Figs. \ref{coh} (c) and (d) show that the evolution of system has a similar
behavior as that of $\beta =\pi /2$. However, due to the coherence of the
initial local state, the oscillation in each envelopment disappears.
Furthermore, the maximum of transition probability is $0.135$ and $0.139$
for atom and photon at time $51.84$ and $51.73$ respectively, which are
about half of those for $\beta =\pi /2$.
\begin{figure}[tbp]
\begin{center}
\includegraphics[width=0.38\textwidth]{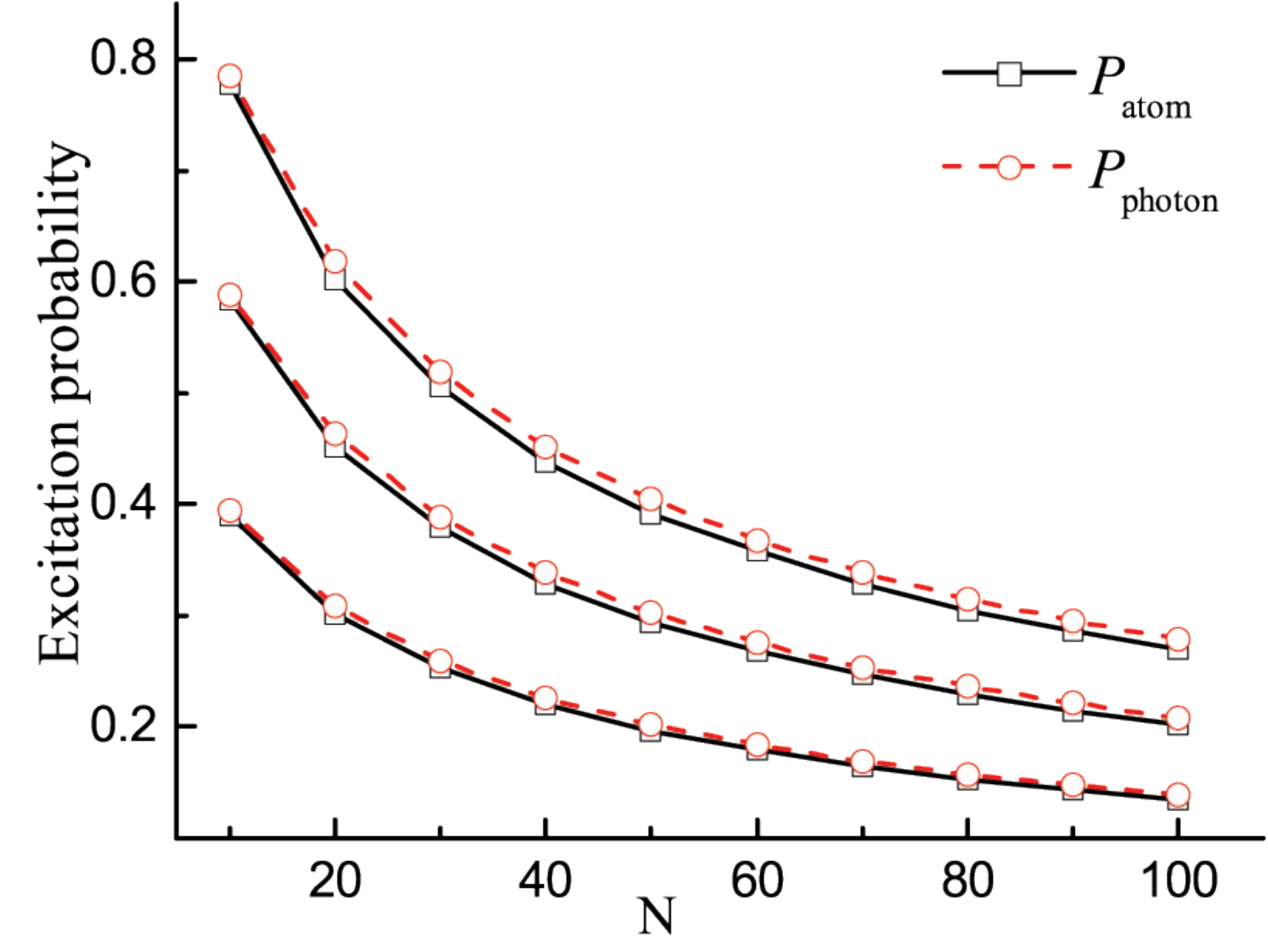}
\end{center}
\caption{(Color online) The binary transmission probabilities verse the
number of cavities with different $\protect\beta $. From top to bottom, $%
\protect\beta =0$, $\protect\beta =\protect\pi /6$, and $\protect\beta =%
\protect\pi /4$, with the system parameters are $\Delta =0$, $\protect%
\lambda =10$, and $\protect\xi =1$.}
\label{fig2}
\end{figure}

The maximum of transmission probabilities for various cavity lengths with
different $\beta $ is numerically evaluated. Fig. \ref{fig2} shows that the
entangled state located in the first cavity does not provide any benefit to
the transmission probability, and the transition probability of atomic
channel is slightly less than that of photonic channel for various values of
$\beta $. Also, the maximum transmission probabilities decrease with the
number of cavities $N$.
\begin{figure}[t]
\begin{center}
\includegraphics[scale=0.28]{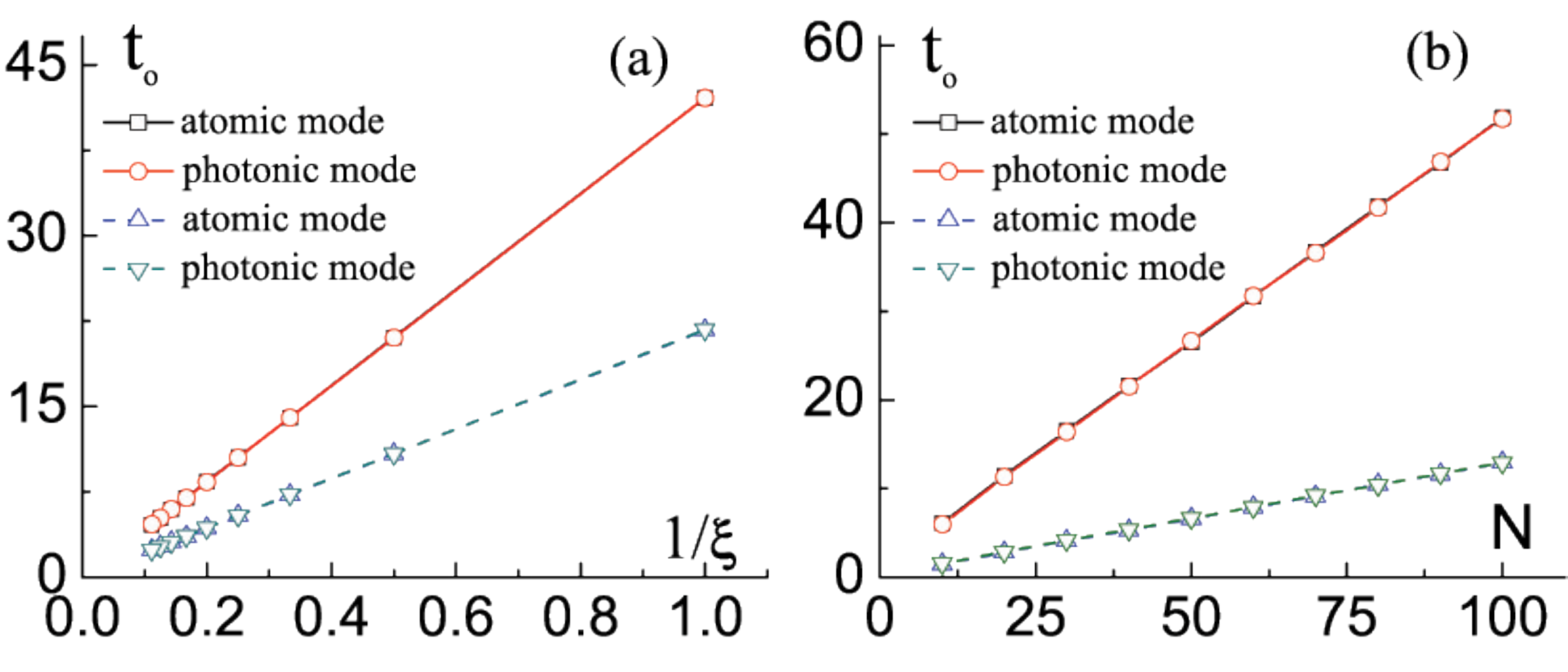}
\end{center}
\caption{(Color online) $\ $On resonance, the optimal time $t_{o}$ of binary
transmission as a function of (a) $\protect\xi $, from top to bottom, $N=80$
and $N=40$ with $\protect\lambda =80$ and $\protect\beta =0$; (b) $N$, from
top to bottom, $\protect\xi =1$ and $\protect\lambda =10$, $\protect\xi =4$
and $\protect\lambda =40$ with $\protect\beta =0$.}
\label{time}
\end{figure}

The binary-channel optimal time $t_{o}$ can be explored when the
transmission probability achieves its maximum. It is found that the time $%
t_{o}$ is not sensitive to the cavity-atom coupling $\lambda $ in the
strongly coupling limit, while it is highly dependent on the number of the
cavities $N$ and intercavity hopping amplitude $\xi $. As shown in Fig. \ref%
{time} (a), $t_{o}$ of atomic and photonic modes is plotted with respect to $%
1/\xi $ for $N=40$ and $N=80$ with $\lambda =80$ when two photons are
initially located in the first cavity. The atomic optimal time is nearly
identical to that of the photonic mode. A linear dependence of both $t_{o}$
on $1/\xi $ is depicted, and the ratio increases with respect to the length
of the cavity array $N$, as is also confirmed in Fig. \ref{time} (b). The
farther the distance is, the more time the propagation needs. The time $%
t_{o} $ scales linearly with $N$ for atomic and photonic modes. Therefore,
the optimal time satisfies $t_{o}\sim N/\xi $. Thus, the optimal time can be
tuned by controlling intercavity hopping strength and system size.

Finally, we give a comparison between this scheme and that based on the
Jaynes-Cummings-Hubbard model \cite{mak}. We note that in contrast to
single-photon JC coupling $g$ used in that model, the two-photon coupling
strength $\lambda \simeq g^{2}/\delta \sim 0.1g$ (assuming $g_{1}=g_{2}=g$).
To preserve the strong coupling condition, $\lambda =10\xi $ for instance,
the speed of the quantum transfer will reduce to $2\xi \simeq 0.02g$
considering the photon's bounding effect, which is about $1/5$ of that of
the Jaynes-Cummings-Hubbard model. With the same parameters used in Figs. %
\ref{coh} (c) and (d), the maximum excitation probabilities of atom and
field are $0.139$ and $0.141$ at time $10.41$ and $10.40$ \cite{mak}. It is
clear that the weakening of the transmission fidelity in our scheme is less
than $0.01$, which is nearly neglectable. However, due to the long lifetime
of metastable states, the distance of the propagation can obviously increase
during the lifetime of atom. For instance, the spontaneous emission
lifetimes of the $D_{5/2}$ metastable state and the $P_{1/2}$ level of $^{40}
$\textrm{Ca}$^{+}$ ion trapped in high finesse optical cavity are about $1s$
\cite{blatt04} and $7ns $ \cite{buzek01} respectively, which leads to an
increase of propagation distance with a seven order of magnitude when the
metastable state is exploited in the scheme.

\section{STAGGERED HOPPING}

Compared with the uniform hopping, a parabolic hopping was firstly advanced
in theory to achieve the perfect quantum transfer, and the dynamic behavior
of such kind of hopping is elaborately investigated in 1D
Jaynes-Cummings-Hubbard model \cite{mak}. Recently, there is another kind of
hopping \cite{ciccarello} called staggered hopping draws much interest. Some
models with the staggered next-nearest coupling have been widely studied in
condensed matter physics. For instance, a Peierls distorted chain \cite{pei}
can be used as a data bus \cite{huo08} to transfer the information. In this
section, we consider the staggered pattern of hopping strengths, i.e., the
hopping strength of the next-nearest cavities are different between
odd-to-even and even-to-odd, which is controlled by the distortion of the
hopping strength $\kappa $. To solve this model, we exploited the same
process dealt with Eq. (\ref{mainHamiltonian}), which first diagonalize the
auxiliary Hilbert space spanned by the site of the cavity $\{\left\vert
M\right\rangle \}$. Here the matrix element of the adjacency matrix $A$ in $%
\mathcal{H}_{\mathrm{res}}$ is rewritten as $A_{ji}=A_{ij}=[1-\kappa
(-1)^{i}]\delta _{i+1,j}$, $(i\leqslant j)$. Here, we assume that the number
of the cavities is odd, and hence the numbers of strong bonds and weak bonds
are equal. The eigenvectors of this adjacency matrix are given as

\begin{align}
\left\vert o\right\rangle & =\frac{2}{\kappa -1}\sqrt{\frac{\kappa }{\tau
^{N+1}-1}}\sum\limits_{M=1}^{\frac{N+1}{2}}\tau ^{M-1}\left\vert
2M-1\right\rangle ,  \label{eqo} \\
\left\vert m\right\rangle _{\pm }& =\sqrt{\frac{2}{N+1}}\bigg[%
\sum\limits_{M=1}^{\frac{N-1}{2}}\sin \left( \frac{2mM\pi }{N+1}\right)
\left\vert 2M\right\rangle  \notag \\
& \pm \sum\limits_{M=1}^{\frac{N+1}{2}}\sin \left( \frac{2mM\pi }{N+1}%
+\theta _{m}\right) \left\vert 2M-1\right\rangle \bigg],\text{ }  \label{eqm}
\end{align}%
with corresponding eigenvalues
\begin{eqnarray}
\varepsilon_{o} &=&0, \\
\varepsilon_{m_{+}} &=&-\varepsilon _{m_{-}}=2\sqrt{\cos ^{2}\frac{m\pi }{%
N+1}+\kappa ^{2}\sin ^{2}\frac{m\pi }{N+1}},
\end{eqnarray}%
where the bond alternation parameter $\tau =\frac{\kappa +1}{\kappa -1}$ and
the angle $\theta _{m}$ satisfies
\begin{equation}
e^{i\theta _{m}}=\frac{1-\kappa }{\varepsilon _{m_{+}}}\left( e^{-i\frac{%
2m\pi }{N+1}}-\tau \right) .\text{ }m=1,2,...,\frac{N-1}{2}
\end{equation}%
Now we investigate the dynamical behavior of the system under this staggered
coupling. In the $\upsilon $th block, the Hamiltonian is given as

\begin{equation}
H_{\mathrm{res}}(\upsilon )=\left(
\begin{array}{cc}
\frac{\Delta }{2} & \sqrt{2}\lambda \\
\sqrt{2}\lambda & -\frac{\Delta }{2}+2\xi \varepsilon _{\upsilon }%
\end{array}%
\right) ,\text{ }\upsilon =o,m_{\pm }
\end{equation}%
The eigenvectors of the $\upsilon $th block of the Hamiltonian are%
\begin{equation}
\left\vert \upsilon ,\pm \right\rangle =\left\vert \upsilon \right\rangle
\otimes \left\vert \pm _{\upsilon }\right\rangle =\left\vert \upsilon
\right\rangle \otimes \frac{\left( \Delta -2E_{\mp }^{\upsilon }\right)
\left\vert e,0\right\rangle +2\sqrt{2}\lambda \left\vert g,2\right\rangle }{%
\sqrt{\left( \Delta -2E_{\mp }^{\upsilon }\right) ^{2}+8\lambda ^{2}}},
\label{block}
\end{equation}%
with the corresponding eigenenergy%
\begin{equation}
E_{\pm }^{\upsilon }=\xi \varepsilon _{\upsilon }\pm \sqrt{\left( \frac{%
\Delta }{2}-\xi \varepsilon _{\upsilon }\right) ^{2}+2\lambda ^{2}}.
\end{equation}%
\begin{figure}[tbp]
\centering\includegraphics[scale=0.3]{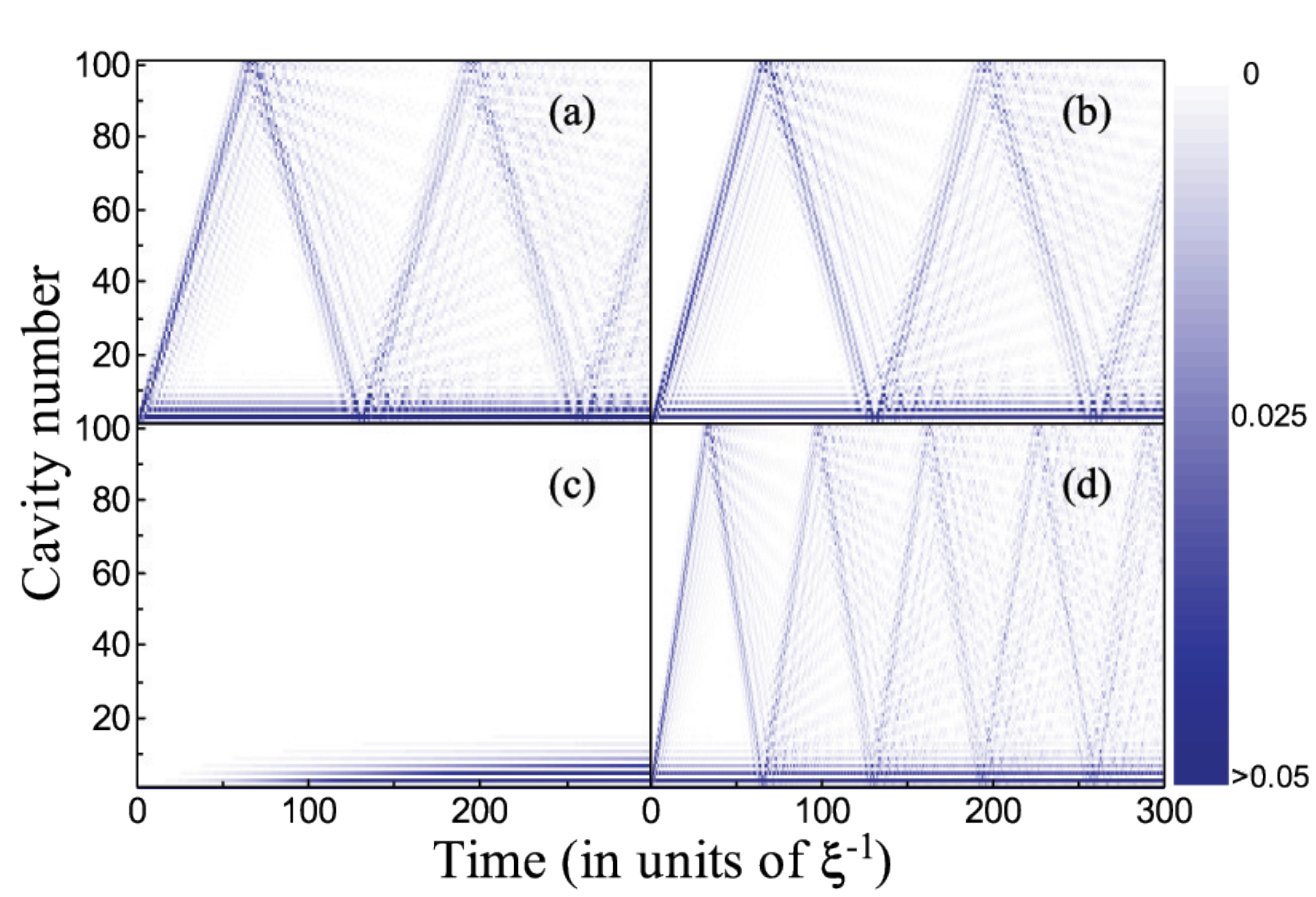} \caption{(Color
online) The probability of the certain cavity as a function of time
with the staggered coupling. The upper two plots show the population
of atom (a) and photon (b) in the strong coupling regime ($\protect\lambda %
=200\protect\xi $), and the lower two plots are the population of atom (c)
and photon (d) in the strong hopping regime ($\protect\lambda =\protect\xi %
/200$). The parameters are $\protect\beta =\protect\pi /4$, $\Delta =0$, $%
N=101$, $\protect\kappa =-0.2$.}
\label{kappatwo}
\end{figure}

\begin{figure}[tbp]
\centering\includegraphics[scale=0.3]{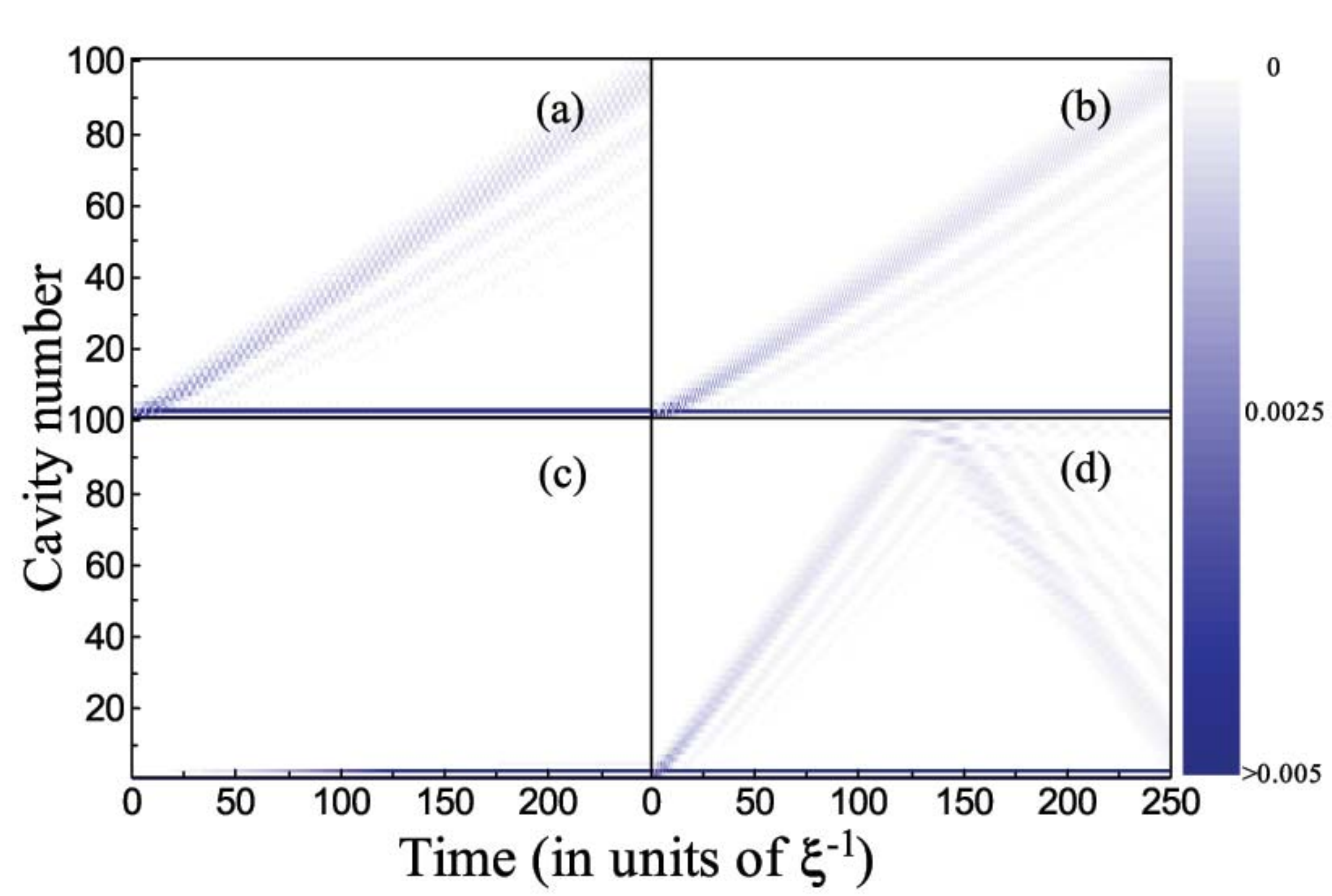}
\caption{(Color online) The probability of the certain cavity as a
function of time with the staggered coupling. The upper two plots
show the population
of atom (a) and photon (b) in the strong coupling regime ($\protect\lambda %
=200\protect\xi $), and the lower two plots are the population of atom (c)
and photon (d) in the strong hopping regime ($\protect\lambda =\protect\xi %
/200$), with the system parameters are $\protect\beta =\protect\pi /4$, $%
\Delta =0$, $N=101$, and $\protect\kappa =-0.8$.}
\label{kappaeight}
\end{figure}

For the initial entangled state Eq. (\ref{ini}) in the first cavity, the
excitation populations of the atom and field in the $s$th cavity after time $%
t$ are%
\begin{eqnarray}
P_{\mathrm{atom,s}} &=&|\sum\limits_{\upsilon ,m}c_{\upsilon }c_{\upsilon
}^{\prime }[\chi ^{-}\left( \alpha _{\upsilon }\right) \chi ^{+}\left( \beta
-\alpha _{\upsilon }\right) e^{-iE_{+}^{\upsilon }t}  \notag \\
&&+\chi ^{+}(\alpha _{\upsilon })\chi ^{-}\left( \beta -\alpha _{\upsilon
}\right) e^{-iE_{-}^{\upsilon }t}]|^{2}.
\end{eqnarray}%
\begin{eqnarray}
P_{\mathrm{photon,s}} &=&|\sum\limits_{\upsilon ,m}c_{\upsilon }c_{\upsilon
}^{\prime }[-\chi ^{+}(\alpha _{\upsilon })\chi ^{+}\left( \beta -\alpha
_{\upsilon }\right) e^{-iE_{+}^{\upsilon }t}  \notag \\
&&+\chi ^{-}(\alpha _{\upsilon })\chi ^{-}\left( \beta -\alpha _{\upsilon
}\right) e^{-iE_{-}^{\upsilon }t}]|^{2},
\end{eqnarray}%
where $\tan \alpha _{\upsilon }=-\{[\left( \Delta -2E_{+}^{\upsilon }\right)
^{2}+8\lambda ^{2}]/[\left( \Delta -2E_{-}^{\upsilon }\right) ^{2}+8\lambda
^{2}]\}^{1/2}$ and the amplitudes $c_{\upsilon }^{\prime }$ are dependent on
the site of the cavity, when $s$ is odd
\begin{gather}
c_{o}^{\prime }=\frac{2}{\kappa -1}\sqrt{\frac{\kappa }{\tau ^{N+1}-1}}\tau
^{\frac{s-1}{2}},  \label{odd1} \\
c_{m_{+}}^{\prime }(m)=-c_{m_{-}}^{\prime }(m)=\sqrt{\frac{2}{N+1}}\sin %
\left[ \frac{(s+1)m\pi }{N+1}+\theta _{m}\right] ,  \label{odd2}
\end{gather}%
and when $s$ is even%
\begin{eqnarray}
c_{o}^{\prime } &=&0,  \label{even1} \\
c_{m_{+}}^{\prime }(m) &=&c_{m_{-}}^{\prime }(m)=\sqrt{\frac{2}{N+1}}\sin
\left( \frac{sm\pi }{N+1}\right) .\text{ }  \label{even2}
\end{eqnarray}%
The amplitudes of the initial state $c_{\upsilon }$ equal to $c_{\upsilon
}^{\prime }$ with $s=1$.

When the distortion $\kappa \rightarrow 0$, it returns to the uniform
coupling discussed above. Here, the behavior of the system is studied in
both strong coupling and hopping region for different values of $\kappa $.
Figs. \ref{kappatwo} (a) and (b) show the evolution of the excitation along
the array in the strong coupling limit when $\kappa =-0.2$. Compare with the
uniform case, the maximum of excitation probability reduces to $0.08$, and
the speed of the propagation slows down lightly. When $\kappa =-0.8$, Figs. %
\ref{kappaeight} (a) and (b) show that it further reduces the speed and
considerably lessens the excitation probability, and the population is
almost localized in the first cavity. In the strong hopping limit, Fig. \ref%
{kappatwo} (c) shows the atomic population distributes on the first several
cavities, especially the first several odd cavities, while the amplitude of
the even sites vanishes. For the photonic mode, the speed is double
referring to the strong coupling and the maximum of excitation probability
also reaches $0.08$ (see Fig. \ref{kappatwo} (d)). When the hopping
dominates, as the distortion increases to $-0.8$, only the photon transfers
along the array with the maximum excitation probability $0.0026$ (see Figs. %
\ref{kappaeight} (c) and (d)).

\section{MULTIQUBIT ENCODING}

It is well known that the highest fidelity for classical transmission of a
quantum state is $2/3$ \cite{horodecki}. However, Fig. \ref{fig2} implies
that even in the case of $\beta =0$, the transmission fidelity exceeds $2/3$
in a very short distance. Therefore, the further improvement of the transfer
probability of the state is necessary. In spin system, there are several
methods to improve the fidelity of state transfer, such as modulating the
couplings \cite{chr}, or turning on/off the coupling on demand \cite{ross09}%
. Alternatively, one can improve the communication by encoding the
information in the multiple spins \cite{linden2004,zou,bish} without
engineering the coupling.

In Ref. \cite{mak}, a Gaussian wave packet is exploited as the initial
state. However, encoding all the qubits through the coupled cavity array is
difficult to achieve in reality. Here, an alternative $k$-qubit encoding is
used, which only needs limited number of qubits to be encoded and decoded by
the sender and receiver. In Sec. III, we have shown that the excitation
populations are almost trapped in the first several cavities either in
strong coupling or strong hopping. So we will focus the quantum transfer in
the uniform coupling discussed in Sec. II rather than the staggered hopping,
especially in the strong coupling regime. If a $k$-qubit encoding is used as
the initial state $\left\vert \varphi (t=0)\right\rangle
_{k}=\sum\nolimits_{\nu =0}^{k-1}\left( -1\right) ^{\nu }\left\vert M=2\nu
+1\right\rangle \otimes \left\vert e,0\right\rangle /\sqrt{k}$, the maximum
value can be achieved for $k=2$, which means that one cannot get a higher
transmitting probability even using more qubits to encode. To obtain a
higher transmitting probability, the decoding process at the another site of
the cavities must be considered. The excitation state of the first cavity
would ideally propagate to the $\left[ N-2\left( r-1\right) \right] $th site
of the cavities, while the excitation state of the last cavity of the
encoding would ideally propagate to the $N$th site of the cavities, i.e., $%
\left\vert \varphi _{\mathrm{idea}}\right\rangle _{\mathrm{atom}%
}=\sum\nolimits_{q=0}^{r-1}\left( -1\right) ^{q}\left\vert
M=M_{q}\right\rangle \otimes \left\vert e,0\right\rangle /\sqrt{r}$ and $%
\left\vert \varphi _{\mathrm{idea}}\right\rangle _{\mathrm{photon}%
}=\sum\nolimits_{q=0}^{r-1}\left( -1\right) ^{q}\left\vert
M=M_{q}\right\rangle \otimes \left\vert g,2\right\rangle /\sqrt{r}$, where $%
M_{q}=N-2\left( r-1\right) +2q$. Thus the probabilities of excitation can be
calculated as%
\begin{equation*}
P_{\mathrm{atom/photon}}=\left\vert \frac{1}{\sqrt{r}}\sum\limits_{m=1}^{N}%
\sum\limits_{q=0}^{r-1}\mu _{m,q}^{\pm }\right\vert ^{2},
\end{equation*}%
where $\mu _{m,q}^{\pm }=(-1)^{q}[\tilde{f}_{M_{q},m}^{\pm }\chi ^{-}\left(
\alpha _{m}\right) \pm \tilde{f}_{M_{q},m}^{\mp }\chi ^{+}\left( \alpha
_{m}\right) ]$, with the probability amplitudes%
\begin{eqnarray}
\tilde{f}_{M,m}^{\pm }\left( t\right) &=&\frac{2}{\left( N+1\right) \sqrt{k}}%
\sum\limits_{\nu =0}^{k-1}\left( -1\right) ^{\nu }e^{-iE_{\pm }^{m}t}\chi
^{\mp }\left( \alpha _{m}\right)  \notag \\
&&\times \sin \left( \frac{mM\pi }{N+1}\right) \sin \left[ \frac{\left( 2\nu
+1\right) m\pi }{N+1}\right] ,
\end{eqnarray}%
\begin{figure}[t]
\begin{center}
\includegraphics[width=0.44\textwidth]{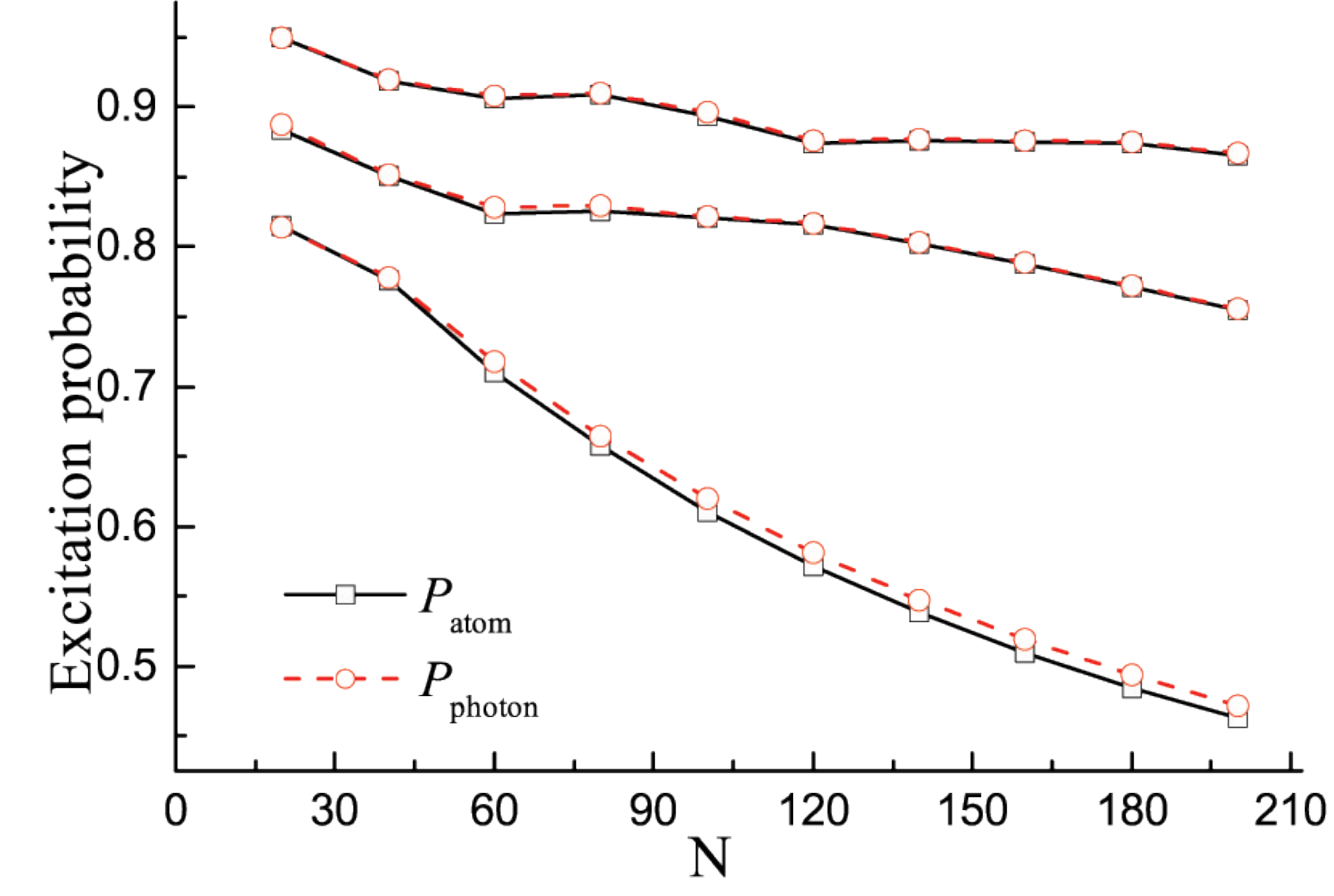}
\end{center}
\caption{(Color online) The binary transmission probabilities verse the
number of cavities with different $k$. From top to bottom, $k=8$, $k=4$, and
$k=2$, with the system parameters are $\Delta =0$, $\protect\lambda =10$,
and $\protect\xi =1$. }
\label{kqubit}
\end{figure}
When the number of encoding qubits equals to that of decoding qubits, the
maximum values of the excitation probabilities can be obtained. In Fig. \ref%
{kqubit}, the maximum transfer probability is plotted as a function of $N$
for the class of encoding states. The curves show that the propagation
fidelity increases rapidly as the number of encoding qubits $k$ increases.
Clearly, the transmission probability decreases with respect to the cavity
number $N$, and the decay becomes slower by encoding and decoding more
qubits. The difference of excitation probabilities between atomic and
photonic mode decreases as the number of encoding qubits increases. As is
displayed in Fig. \ref{kqubit}, the maximum probabilities of the binary
transmission are higher than $0.86$ for $k=8$, even up to the $200$ cavities.

\section{CONCLUSION}

To conclude, the dynamics of a cavity array is studied when each cavity
contains a three-level atom. Adiabatically eliminating the intermediate
state, the individual cavity can be described by the two-photon JC model
with a metastable state and a ground state. By exploiting the metastable
state, the transfer distance can be much longer than that of
Jaynes-Cummings-Hubbard model at the expense of longer transmission times.
When the adjacent cavities are coupled by two-photon hopping, the whole
system Hamiltonian can be exactly diagonalized in the subspace of
single-atom excitation. For the uniform and staggered intercavity hopping,
we analyze the dynamics of the system in distinct regimes of parameters.
With the staggered hopping, the excitation population trends to be localized
in the first several cavities and slows down the transfer speed. This
phenomenon becomes more distinct as the distortion of the hopping strength
increases. Compared with the staggered hopping, the uniform hopping is more
suitable for the quantum state transfer. In the strong hopping limit, only
the photon is delocalized on the whole array of the coupled cavity with
double speed. While in the strong coupling limit, the initial state in the
first cavity plays an important role in the transmission fidelity for the
binary transmission. Due to the strong coupling between atom and quantum
field in individual cavity, the difference of maximum probabilities between
the binary transmission is not very large. Finally, it is shown that a class
of initial encodings and final decoding process used in spin system can also
greatly improve the performance of binary transmission. The analysis of the
dynamics in the high-Q optical cavities can also be exploited in analogous
systems, such as circuit quantum electrodynamics and coupled photonic
crystal cavities. The results of this work provide a step for studying the
quantum state transfer in coupled cavities.

\section{ACKNOWLEDGMENTS}

This work is supported by the Specialized Research Fund for the Doctoral
Program of Higher Education (Grant No. 20103201120002), Special Funds of the
National Natural Science Foundation of China (Grant No. 11047168), the
National Natural Science Foundation of China (Grant Nos. 11074184 and
11004144), the Natural Science Foundation of Jiangsu Province (Grant No.
10KJB140010), and the PAPD.

\end{document}